\documentclass[a4paper,fleqn,usenatbib]{mnras}
\usepackage{newtxtext,newtxmath}

\usepackage[T1]{fontenc}
\usepackage{ae,aecompl}

\usepackage{graphicx}	
\usepackage{amsmath}	
\usepackage{amssymb}	

\newcommand{\numax}{\nu_{\mathrm{max}}}

\title[Solar cycle variation of $\numax$]{Solar cycle variation of $\numax$ in helioseismic data and its implications for asteroseismology}

\author[R. Howe et~al.]{Rachel Howe,$^{1,2}$\thanks{E-mail: r.howe@bham.ac.uk    (RH)}
William J. Chaplin,$^{1,2}$
Sarbani Basu,$^{3}$
Warrick H. Ball,$^{1,2}$ \newauthor
Guy R. Davies,$^{1,2}$
Yvonne Elsworth,$^{1,2}$ 
Steven J. Hale,$^{1,2}$
Andrea Miglio,$^{1,2}$ \newauthor
Martin Bo Nielsen,$^{1,2,4}$
and Lucas S. Viani$^{3}$
\\
$^{1}$School of Physics and Astronomy,
  University of Birmingham, Birmingham, B15 2TT, United
  Kingdom\\ $^{2}$Stellar Astrophysics Centre (SAC), Department of
  Physics and Astronomy, Aarhus University,\\ Ny Munkegade 120, DK-8000
  Aarhus C, Denmark\\ $^{3}$ Department of Astronomy, Yale University, P.O. Box 208101, New Haven, CT 06520-8101, USA\\ $^4$ Center for Space Science, NYUAD Institute, New York University Abu Dhabi, PO Box 129188, Abu
Dhabi, United Arab Emirates}

\date{Accepted XXX. Received YYY; in original form ZZZ}

\pubyear{2019}

\begin{document}
\label{firstpage}
\pagerange{\pageref{firstpage}--\pageref{lastpage}}
\maketitle

\begin{abstract}
  The frequency, $\numax$, at which the envelope of pulsation power peaks for solar-like oscillators is an important quantity in asteroseismology. We measure $\numax$ for the Sun using 25 years of Sun-as-a-Star Doppler velocity observations with the Birmingham Solar-Oscillations Network (BiSON), by fitting a simple model to binned power spectra of the data. We also apply the fit to Sun-as-a-Star Doppler velocity data from GONG and GOLF, and photometry data from VIRGO/SPM on the ESA/NASA SOHO spacecraft. We discover a weak but nevertheless significant positive correlation of the solar $\numax$ with solar activity. The uncovered shift between low and high activity, of $\simeq 25\,\rm \mu Hz$, translates to an uncertainty of 0.8\,per cent in radius and 2.4\,per cent in mass, based on direct use of asteroseismic scaling relations calibrated to the Sun. The mean $\numax$ in the different datasets is also clearly offset in frequency. Our results flag the need for caution when using $\numax$ in asteroseismology. 
  
\end{abstract}




\section{Introduction}
\label{sec:intro}

In the Sun and other solar-like oscillators, the frequency at which the envelope of the pulsation spectrum has its maximum in power is known as $\numax$ \citep{kjeldsen95}. The quantity plays a role in ensemble asteroseismology, as via scaling relations \citep[e.g.][]{chaplin13} it can be used to help constrain fundamental stellar properties, even though it is set by the characteristics of the outer layers of the star where the modes are excited and damped \citep{belkacem11, belkacem13}. It was \citet{brown91} who first suggested explicitly that $\numax$ might be related to the photospheric acoustic cut-off frequency, which in turn suggested a scaling with surface gravity, $g$, and effective temperature, $T_{\rm eff}$, of the form 
 \begin{equation}
 \numax \propto g T_{\rm eff}^{-1/2} \propto MR^{-2} T_{\rm eff}^{-1/2},
 \end{equation}
with $M$ the mass and $R$ the radius of the star.  The use of $\numax$ to constrain stellar properties rests not only on the level of accuracy of this scaling \citep[e.g., see][]{coelho15, viani17}, but also on having a robust reference to calibrate the relation. The solar $\numax$ is usually adopted as the reference. Recent estimates of it in the literature cover the range 3080 to $3160\,\rm \mu Hz$. This spread translates to a difference of about 2.5\,per cent in the radius and 8\,per cent in the mass for a given star, based on direct use of scaling relations \citep[e.g.][]{chaplin13} calibrated to the Sun. 

Many estimates of the solar $\numax$ come from using photometric Sun-as-a-star data \citep{kallinger10, huber11, mosser13, kallinger14} collected by the VIRGO/SPM instrument on the ESA/NASA SOHO spacecraft \citep{froehlich95}. Another photometric estimate used reflected sunlight from Neptune captured by the NASA \emph{Kepler} telescope during its extended K2 mission phase \citep{gaulme16}. Others have instead estimated $\numax$ using data in Doppler velocity, e.g., from the Global Oscillation Network Group (GONG) telescopes \citep{kiefer18}; or from the Stellar Observations Network Group (SONG) spectrograph fed with scattered sunlight \citep{andersen19}.

There are several factors that contribute to the spread in the reported values. Differences in methodology may play a role, e.g., the parametric model used to fit the data. For example, fits to  stars and solar data typically assume a Gaussian envelope for the pulsation spectrum \citep[e.g.][]{lund17}; however, the solar envelope is clearly asymmetric \citep{kiefer18}. 

Differences related to the data are undoubtedly important, beginning with how the oscillations are observed. Doppler results tend to give a higher $\numax$ than do photometric observations. There are also differences between different Doppler datasets. Doppler observations are sensitive to perturbations at different heights in the stellar atmosphere, depending on the Fraunhofer line (or lines) used. The locations of the outer boundaries of the mode cavities lie beneath the photosphere, and depend on mode frequency. Perturbations due to the modes are therefore evanescent in the photosphere, with the observed signal suffering frequency dependent attenuation with increasing height. This has the potential to affect the observed $\numax$. Also of note is that the ratio of the amplitudes of signals due to oscillations and granulation is much lower in photometry than in Doppler velocity. This means that for photometric data, the background power spectral density is dominated in frequency by the granulation, and the oscillations are usually observed at a significantly lower signal-to-noise ratio than in Doppler velocity. Most results come from Sun-as-a-star data, which are sensitive to modes of low angular degree, $l$; but one recent estimate has been made using a much wider set of modes ($2 \le l \le 150$; see \citealt{kiefer18}). Finally, even though data from the same instrument may have been used, the selections are not usually contemporaneous. It is on this last point that we focus here.

While most stellar observations cover relatively short time intervals, for the Sun we have more than twenty years of high-quality observations from multiple instruments, in multiple observables. It is therefore of interest to see how precisely we can in fact measure the solar $\numax$ over long periods of time using different instruments, as this has implications for the use of the quantity in scaling relations, and to see whether there is any intrinsic variability related to the solar cycle.  That is the objective of this paper.

\section{Data and Methods}
\label{sec:data}

Our primary dataset consists of Sun-as-a-star Doppler velocity observations from the Birmingham Solar-Oscillations Network \citep[BiSON; ][]{chaplin96,hale16}. The data we used cover the period 1995 to early 2018, or nearly two full solar cycles. For comparison, we also considered velocity data from the Global Oscillation Network Group \citep[GONG;][]{harvey96} and from the Global Oscillations at Low Frequency (GOLF) instrument \citep{gabriel95} on SOHO; and photometry data from the Red channel of the VIRGO/SPM instrument \citep{froehlich97}, also on SOHO. For GONG we used the $l=0$ spherical harmonic time series provided by the GONG project\footnote{Available from \url{gong.nso.edu}}. The GONG data have been treated with a first-difference filter; to correct for the effects of this on the spectral power we divide the spectrum by a factor of $4\sin^2(\pi\nu/2\nu_{{\mathrm{Nyq}}})$, where $\nu_{{\mathrm{Nyq}}}$ is the Nyquist frequency -- $8333\,\rm \mu Hz$ for the 60-second cadence of GONG. For VIRGO/SPM, we used up-to-date level-1 data\footnote{\url{SOHO.nascom.nasa.gov/data/data.html}} from the Red channel. The 22-year GOLF dataset comes from a new calibration and is an average of signals from the PM1 and PM2 detectors \citep{appourchaux18}.

The procedure for measuring $\numax$ over time was as follows. The time series of observations were divided into overlapping segments -- each of length 1\,year, with start-times offset by 3\,months -- and a Fourier power spectrum was computed for each segment. Each spectrum was then averaged over a number of frequency bins of width 135\,$\mu$Hz, corresponding to the separation between modes of the same degree and adjacent radial order, the so-called large frequency separation $\Delta\nu$. Binning gives data that have an underlying smooth trend in frequency and statistics that tend to Gaussian.

We used the \texttt{lmfit} python package \citep{LMFIT2018} to perform a non-linear least-squares fit of a custom model to each binned power spectrum. Taking inspiration from \citet{kiefer18}, the model consists of an asymmetric pseudo-Voigt profile \citep{STANCIK200866}. There is also a frequency-dependent background term, and a constant background offset. The oscillation envelope profile takes the form
 \begin{equation}
 P(\nu)= f \times P_L(\nu) + [1-f] \times P_G(\nu),
 \end{equation}  
where $\nu$ is frequency, $P_L(\nu)$ and $P_G(\nu)$ are respectively the Lorentzian and Gaussian parts of the profile, and $f$ is the factor governing the balance between the Lorentzian and Gaussian contributions. Specifically, we use
 \begin{equation}
 X=[(\nu-\numax)/\Gamma(\nu)]^2,
 \end{equation}
where 
 \begin{equation}
 \Gamma(\nu)=2\Gamma_0/[1+\exp^{[a(\nu-\numax)]}]
 \end{equation}
describes the frequency dependence of the width, with $\Gamma=\Gamma_0$ at $\nu=\numax$ and $a$ being an asymmetry term. The Lorentzian term is 
 \begin{equation}
 {P_L}(\nu)={\frac{2H}{(1+4X)\pi\Gamma(\nu)}},
 \end{equation}  
and the Gaussian term is
 \begin{equation}
 {P_G}(\nu)={\frac{H\sqrt{4\ln(2)}}{\pi\Gamma(\nu)\exp^{[4X\ln(2)]}}},
 \end{equation}
where $H$ governs the height of the oscillation power envelope.

A weakness of this parametrisation is that if both $\numax$ and $a$ are allowed to vary independently the fit results for these two parameters are highly correlated. For the final fits we therefore selected a fixed value of $a$, by repeating the fits for a range of $a$ and selecting the one that gave the lowest overall $\chi^2$ for the whole dataset. For the background we tried both a Lorentzian model centred on zero frequency, and a non-parametric background given by smoothing with a median filter.

\begin{figure}
	\centering
	\includegraphics[width=0.45\textwidth]{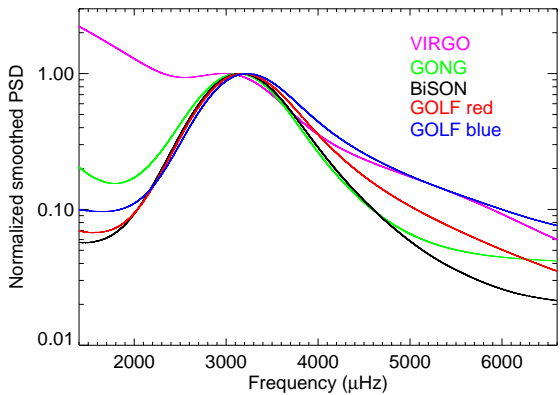}
	\caption{Mean of the bin-averaged power spectra from each full dataset, normalized to show the same power spectral density at the peak of the five-minute envelope (arbitrarily scaled to unity on the ordinate).}
	\label{fig:figspecs}
\end{figure}

Fig.~\ref{fig:figspecs} shows the mean of the bin-averaged power spectra from each full dataset, all normalized to show the same power spectral density at the peak of the five-minute envelope (arbitrarily scaled to unity on the plot). The figure shows clearly differences in the shape of the observed power spectra. The final fits adopted fixed values of the asymmetry parameter of $a=-0.5$ (BiSON), $-0.3$ (GONG) and $-0.7$ (VIRGO). Note a negative asymmetry implies more power on the high-frequency side of the oscillation envelope. Two mean spectra are shown for GOLF: one made from data collected after 1998 September and before 2002 November, when in our final fits $a$ was fixed at $-1.2$; and another made from data collected at earlier and later dates, when it was fixed at $a=-1.9$. This reflected a change in the instrument's observing mode, from detecting Doppler shifts before 1998 September in the blue wing of the Sodium doublet at 589\,nm; to then detecting them in the red wing up to 2002 November; and finally to detecting them back in the blue wing thereafter. We found that the free parameter $f$ converged on values typically around 20\,\% or lower for VIRGO, in the range 30\,\% to 40\,\% for BiSON and GONG, around 60\,\% for GOLF red wing, and even higher for GOLF blue wing.


The statistics of the bin-averaged spectrum mean that the uncertainties in each power average are highest near $\numax$, where the modes are most prominent. That is because individual bins in each $\Delta\nu$-wide segment that contribute to the re-binned averages will span a considerable range in power, from high-power spectral densities across individual modes down to lower-power spectral densities between modes. An error-weighted fit would then be dominated by the averaged bins in the wings of the power envelope and the background, where the dispersion from individual bins contributing to each average is lower. We therefore choose to fit the logarithmic spectrum, because this gives similar relative uncertainties in every bin. We used a Monte Carlo method to estimate the uncertainties, in which the fit was repeated for 1000 realizations of each spectrum taken from a normal distribution of width $\sigma_P(\nu)$ centered on the observed value $P(\nu)$ at each frequency bin, and the uncertainty was taken to be the standard deviation of the resulting estimated parameters.

\section{Results}
\label{sec:results}


\begin{figure*}
	\centering
	\includegraphics[width=0.5\textwidth]{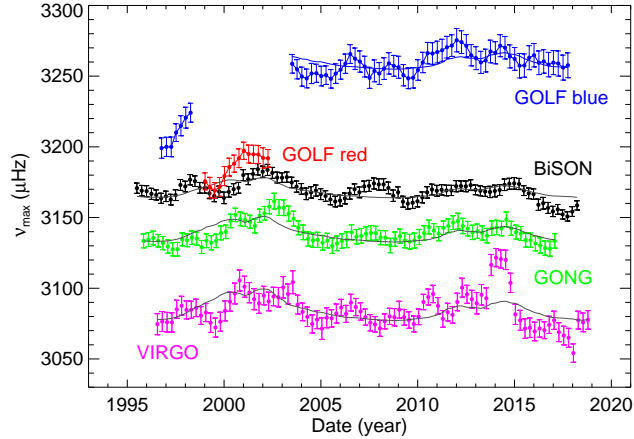}
	\caption{Best-fitting $\numax$ from each dataset as a function of time (coloured points with associated error bars).  The results are for BiSON (black points), GONG (green points), VIRGO (violet points), GOLF blue wing (blue points) and GOLF red wing (red points). The solid lines show the scaled 10.7\,cm radio flux from fits of the linear model defined by Equation~\ref{eq:fitact}.}
	\label{fig:figfits}
\end{figure*}



\begin{figure*}
	\centering
	\includegraphics[width=0.55\textwidth]{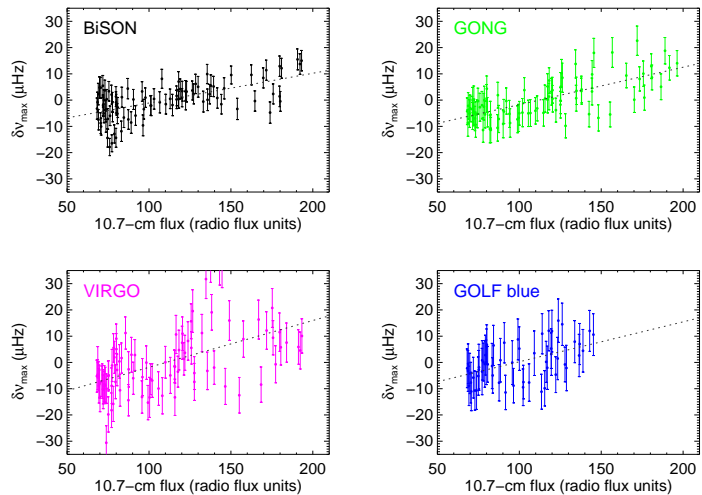}
	\caption{Best-fitting $\numax$ as a function of the 10.7\,cm radio flux, for BiSON (top left-hand panel), GONG (top right-hand panel), VIRGO (bottom left-hand panel) and GOLF blue wing (bottom right-hand panel. The dotted lines show fits of the linear model defined by Equation~\ref{eq:fitact}.}
	\label{fig:figact}
\end{figure*}


We comment first on the variability of $\numax$ over time, which is the focus of the paper. Fig.~\ref{fig:figfits} shows the best-fitting $\numax$ from each dataset as a function of time (coloured points with associated error bars).  The results for BiSON (black points), GONG (green points) and VIRGO (violet points) show a similar pattern of variation, and a significant positive correlation with solar activity.  This correlation is shown clearly in Fig.~\ref{fig:figact}, which plots the best-fitting $\numax$ as a function of the 10.7\,cm radio flux, $F_{\rm 10.7}$ \citep{Tapping2013}, which we adopt as a proxy of global solar activity. The dotted lines in Fig.~\ref{fig:figact} are from fits of a simple linear model of the form:
 \begin{equation}
     \nu_{\rm max}(t) = c_0 + c_1 \times \left(F_{\rm 10.7}(t) - 110 \right).
     \label{eq:fitact}
 \end{equation}
The offset of 110\, radio flux units corresponds to the average 10.7-cm flux over the full period of the data; its introduction to the model means $c_0$ corresponds to $\numax$ at average activity. The solid lines in Fig.~\ref{fig:figfits} show the scaled 10.7\,cm radio flux from this linear model. Table~1 reports the best-fitting coefficients of each fit.


\begin{table}
\begin{center}

\caption{Results of fits of the best-fitting $\numax$ from Fig.~\ref{fig:figfits} to the 10.7-cm radio flux, using the linear model described by Equation~\ref{eq:fitact}.}

\begin{tabular}{lcc}
\hline
Dataset& $c_0$ & $c_1$\\  
       & ($\mu\rm Hz$)& ($\mu \rm Hz$\,RF$^{-1}$)\\  
\hline
    VIRGO& $3085 \pm 4$& $0.18 \pm 0.06$\\
     GONG& $3138 \pm 2$& $0.14 \pm 0.03$\\
    BiSON& $3169 \pm 2$& $0.11 \pm 0.03$\\
GOLF blue& $3261 \pm 4$& $0.15 \pm 0.07$\\
\hline
\end{tabular}
\end{center}
\label{tab:fit}
\end{table}


We also performed the analysis using two independent pipelines to verify the results. One pipeline \citep{nielsen17} used a Markov chain Monte Carlo sampler to fit the raw (unaveraged) power spectra to a model comprising three background terms (two Lorentzians and a flat offset) and a Gaussian for the mode envelope. The other pipeline adopted a very different approach, following \citet{huber09}. In brief, after dividing out the background -- as estimated using a moving-median filter -- we computed the autocorrelation of $675\,\rm \mu Hz$ ($\simeq 5\Delta\nu$) wide ranges of the power spectrum, sliding a frequency window through the full spectrum. We then fitted a Gaussian to the sum over all lags to estimate $\numax$, with the correlation between overlapping segments included in the fit. Both pipelines uncovered the same temporal variations as our main pipeline, showing our results on the time variation of $\numax$ are robust against the choice of fitting model and fitting procedure.

We see a positive shift in $\numax$ of $\simeq 25\,\rm \mu Hz$ between low and high activity. The results for GOLF are however more complicated, and are dominated by the change in observing mode. $\numax$ is around $80\,\rm \mu Hz$ lower in the red-wing data than it is in the blue-wing data. If we separate out the blue-wing data, which cover a much longer period, it too shows a positive correlation with activity.

\citet{barban13} studied the solar $\numax$ in data from GOLF and VIRGO collected between 1996 and 2004 and found that the GOLF $\numax$ appeared to be anti-correlated with the sunspot number but that from VIRGO was not. Based on our results above, it now seems clear that the apparent anti-correlation with activity they reported is actually due to the change in operation from one wing to another (and back again). That \citet{barban13} found no apparent change in the VIRGO $\numax$ is likely due to them having had less data than are available to us now, i.e., the changes we uncover, whilst significant, are nonetheless quite weak.


\begin{figure*}
	\centering
	\includegraphics[width=0.47\textwidth]{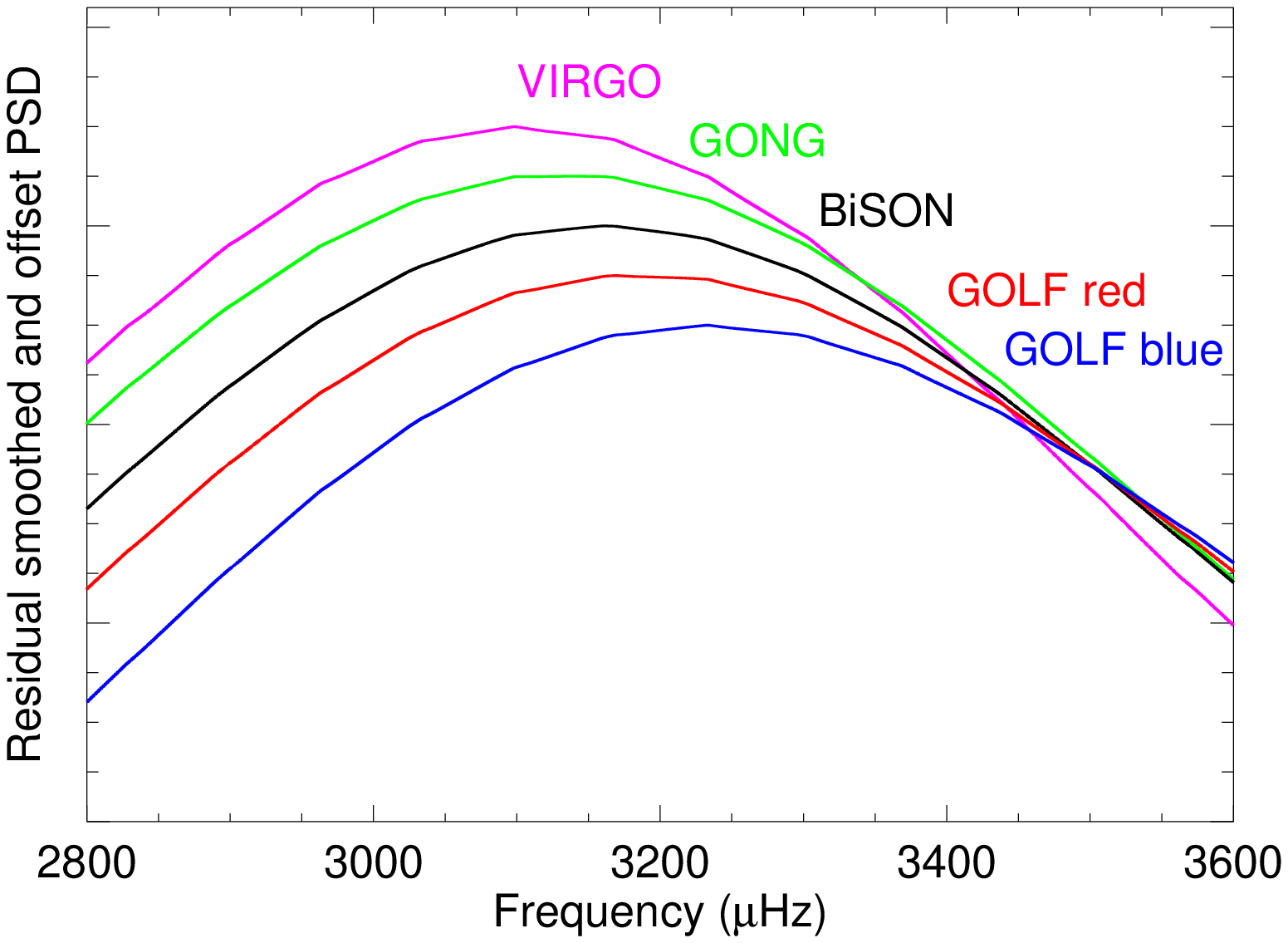}
	\includegraphics[width=0.47\textwidth]{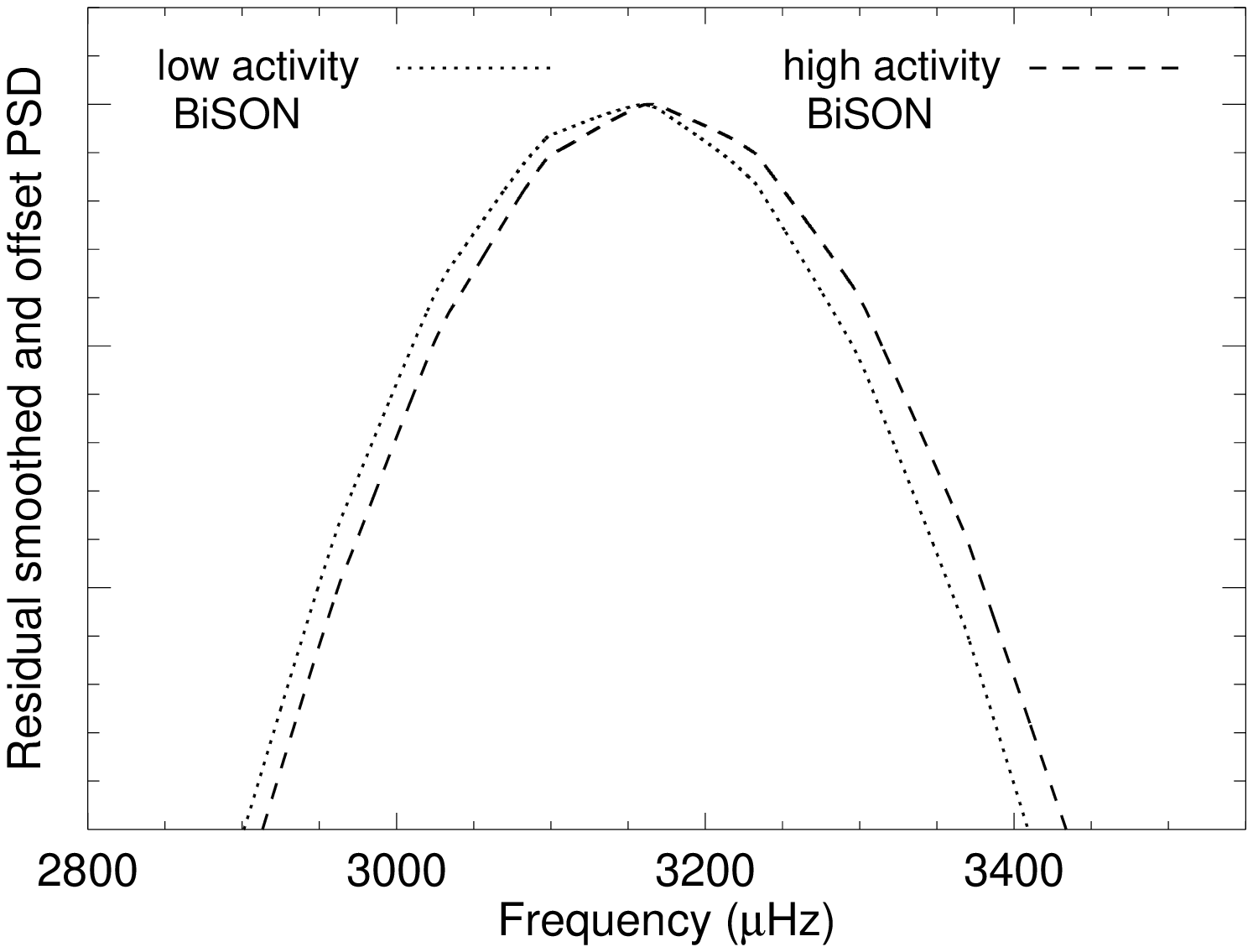}
	\caption{Left-hand panel: Zoom of the mean spectra from Fig.~\ref{fig:figspecs} after a best-fitting background model was removed from each spectrum. The residual spectra have then been offset on the ordinate. Right-hand panel: zoom of mean BiSON spectra from epochs of low and high activity (centered on epochs around 2001 and 2017, respectively).}
	\label{fig:figshift}
\end{figure*}


Our results also show significant differences in the average $\numax$ for the different datasets (e.g. see the $c_0$ estimates in Table~1). The left-hand panel of Fig.~\ref{fig:figshift} shows a zoom of the mean spectra from Fig.~\ref{fig:figspecs} \textsl{after} a mean best-fitting background model was removed from each. The residual spectra have then been offset on the ordinate to show more clearly the differences in $\numax$. The absolute spread is noticeably larger than the activity-related variations. The photometric VIRGO results show the lowest average $\numax$. The right-hand panel shows a zoom of mean BiSON spectra from epochs of low and high activity (centered on epochs around 2001 and 2017, respectively). The cycle-related shift is clearly visible.

\section{Discussion and conclusions}
\label{sec:disc}

We have discovered a weak but nevertheless significant positive correlation of the solar $\numax$ with solar activity. The uncovered shift, of $\simeq 25\,\rm \mu Hz$ between low and high activity, translates to an uncertainty of up to 0.8\,per cent in radius and 2.4\,per cent in mass, based on using the Sun as a calibrator in direct use of the scaling relations. Our result also suggests that we might expect to find variations of $\numax$ in other stars, which would add additional uncertainty to any inferences made using this global asteroseismic parameter (a point we come back to below).

The underlying causes of the variations we have uncovered in the solar $\numax$ must be intimately tied to variations in the power and damping of the modes. There is an extensive literature (see \citealt{howe15}, and references therein) showing that p-mode powers decrease whilst damping rates increase as levels of solar activity rise. The interplay between the relative sizes of these changes with frequency will determine the variation of $\numax$.

As noted earlier, the scaling relation for $\numax$ arose by assuming a proportionate scaling with the photospheric acoustic cut-off frequency. It is therefore intriguing to note that \citet{jimenez11} found a clear positive correlation of the solar cut-off frequency with solar activity. The fractional change they found is about four-times larger than the fractional change we have uncovered in $\numax$.

We also find significant differences in the average $\numax$ given by the various datasets. The ordering of $\numax$  does not respect a simple correlation (or anti-correlation) with the average height above the base of the photosphere at which the observations are effectively made. A correlation with increasing atmospheric height would, moving outwards, suggest (e.g., see \citealt{jimenezreyes07}) an ordering in $\numax$ of VIRGO/SPM, GONG, BiSON, GOLF blue, and finally GOLF red (and a reversed order for an anti-correlation); this is not quite what we see. Nevertheless, that there is such a dramatic change between GOLF red and GOLF blue shows that the way the observations are made matters (see also \citealt{garcia05}). Understanding these absolute differences more clearly will be the focus of future work.

Our results clearly flag the need for caution when using $\numax$ on other stars. They sound a warning over making sure one has a solar reference extracted from data with the same instrumental response as the stellar data, using the same analysis methodology, and for a clearly defined level of solar activity. Since changing levels of activity perturb the solar $\numax$, a desirable reference would be one commensurate with minimal levels of activity (corresponding to minimal impact on $\numax$). But the stellar $\numax$ must then also be compatible with regard to activity levels, which may not always be possible to achieve.

\section*{Acknowledgements}

We would like to thank all those who are, or have been, associated with BiSON, in particular, P. Pall\'e and T. Roca-Cort\'es in Tenerife and E. Rhodes Jr and S. Pinkerton at Mt. Wilson. BiSON is funded by the Science and Technology Facilities Council (STFC). This work utilizes data obtained by the GONG program, managed by the National Solar Observatory, which is operated by AURA, Inc. under a cooperative agreement with the National Science Foundation. The GOLF and VIRGO instruments on board SOHO are a cooperative effort of many individuals to whom we are indebted. SOHO is a project of international collaboration between ESA and NASA. Funding for the Stellar Astrophysics Centre is provided by the Danish National Research Foundation (grant agreement No.: DNRF106). This work made use of the Python packages 
SciPy \citep {scipy}, NumPy \citep{oliphant2006guide}, and Astropy,\footnote{http://www.astropy.org} a community-developed core Python package for Astronomy \citep{astropy:2013, astropy:2018}.

\bibliography{numax} 

\begin{thebibliography}{}
\makeatletter
\relax
\def\mn@urlcharsother{\let\do\@makeother \do\$\do\&\do\#\do\^\do\_\do\%\do\~}
\def\mn@doi{\begingroup\mn@urlcharsother \@ifnextchar [ {\mn@doi@}
  {\mn@doi@[]}}
\def\mn@doi@[#1]#2{\def\@tempa{#1}\ifx\@tempa\@empty \href
  {http://dx.doi.org/#2} {doi:#2}\else \href {http://dx.doi.org/#2} {#1}\fi
  \endgroup}
\def\mn@eprint#1#2{\mn@eprint@#1:#2::\@nil}
\def\mn@eprint@arXiv#1{\href {http://arxiv.org/abs/#1} {{\tt arXiv:#1}}}
\def\mn@eprint@dblp#1{\href {http://dblp.uni-trier.de/rec/bibtex/#1.xml}
  {dblp:#1}}
\def\mn@eprint@#1:#2:#3:#4\@nil{\def\@tempa {#1}\def\@tempb {#2}\def\@tempc
  {#3}\ifx \@tempc \@empty \let \@tempc \@tempb \let \@tempb \@tempa \fi \ifx
  \@tempb \@empty \def\@tempb {arXiv}\fi \@ifundefined
  {mn@eprint@\@tempb}{\@tempb:\@tempc}{\expandafter \expandafter \csname
  mn@eprint@\@tempb\endcsname \expandafter{\@tempc}}}

\bibitem[\protect\citeauthoryear{{Appourchaux}, {Boumier}, {Leibacher}  \&
  {Corbard}}{{Appourchaux} et~al.}{2018}]{appourchaux18}
{Appourchaux} T.,  {Boumier} P.,  {Leibacher} J.~W.,   {Corbard} T.,  2018,
  \mn@doi [\aap] {10.1051/0004-6361/201833535}, \href
  {https://ui.adsabs.harvard.edu/abs/2018A&A...617A.108A} {617, A108}

\bibitem[\protect\citeauthoryear{{Astropy Collaboration} et~al.,}{{Astropy
  Collaboration} et~al.}{2013}]{astropy:2013}
{Astropy Collaboration} et~al., 2013, \mn@doi [\aap]
  {10.1051/0004-6361/201322068}, \href
  {http://adsabs.harvard.edu/abs/2013A%26A...558A..33A} {558, A33}

\bibitem[\protect\citeauthoryear{{Barban}, {Beuret}, {Baudin}, {Belkacem},
  {Goupil}  \& {Samadi}}{{Barban} et~al.}{2013}]{barban13}
{Barban} C.,  {Beuret} M.,  {Baudin} F.,  {Belkacem} K.,  {Goupil} M.~J.,
  {Samadi} R.,  2013, in Journal of Physics Conference Series. p. 012031,
  \mn@doi{10.1088/1742-6596/440/1/012031}

\bibitem[\protect\citeauthoryear{{Belkacem}, {Goupil}, {Dupret}, {Samadi},
  {Baudin}, {Noels}  \& {Mosser}}{{Belkacem} et~al.}{2011}]{belkacem11}
{Belkacem} K.,  {Goupil} M.~J.,  {Dupret} M.~A.,  {Samadi} R.,  {Baudin} F.,
  {Noels} A.,   {Mosser} B.,  2011, \mn@doi [\aap]
  {10.1051/0004-6361/201116490}, \href
  {https://ui.adsabs.harvard.edu/abs/2011A&A...530A.142B} {530, A142}

\bibitem[\protect\citeauthoryear{{Belkacem}, {Samadi}, {Mosser}, {Goupil}  \&
  {Ludwig}}{{Belkacem} et~al.}{2013}]{belkacem13}
{Belkacem} K.,  {Samadi} R.,  {Mosser} B.,  {Goupil} M.~J.,   {Ludwig} H.~G.,
  2013, in {Shibahashi} H.,  {Lynas-Gray} A.~E.,  eds,  Astronomical Society of
  the Pacific Conference Series Vol. 479, Progress in Physics of the Sun and
  Stars: A New Era in Helio- and Asteroseismology. p.~61 (\mn@eprint {arXiv}
  {1307.3132})

\bibitem[\protect\citeauthoryear{{Brown}, {Gilliland}, {Noyes}  \&
  {Ramsey}}{{Brown} et~al.}{1991}]{brown91}
{Brown} T.~M.,  {Gilliland} R.~L.,  {Noyes} R.~W.,   {Ramsey} L.~W.,  1991,
  \mn@doi [\apj] {10.1086/169725}, \href
  {https://ui.adsabs.harvard.edu/abs/1991ApJ...368..599B} {368, 599}

\bibitem[\protect\citeauthoryear{{Chaplin} \& {Miglio}}{{Chaplin} \&
  {Miglio}}{2013}]{chaplin13}
{Chaplin} W.~J.,  {Miglio} A.,  2013, \mn@doi [\araa]
  {10.1146/annurev-astro-082812-140938}, \href
  {https://ui.adsabs.harvard.edu/abs/2013ARA&A..51..353C} {51, 353}

\bibitem[\protect\citeauthoryear{{Chaplin} et~al.,}{{Chaplin}
  et~al.}{1996}]{chaplin96}
{Chaplin} W.~J.,  et~al., 1996, \mn@doi [\solphys] {10.1007/BF00145821}, \href
  {http://adsabs.harvard.edu/abs/1996SoPh..168....1C} {168, 1}

\bibitem[\protect\citeauthoryear{{Coelho}, {Chaplin}, {Basu}, {Serenelli},
  {Miglio}  \& {Reese}}{{Coelho} et~al.}{2015}]{coelho15}
{Coelho} H.~R.,  {Chaplin} W.~J.,  {Basu} S.,  {Serenelli} A.,  {Miglio} A.,
  {Reese} D.~R.,  2015, \mn@doi [\mnras] {10.1093/mnras/stv1175}, \href
  {https://ui.adsabs.harvard.edu/abs/2015MNRAS.451.3011C} {451, 3011}

\bibitem[\protect\citeauthoryear{{Fredslund Andersen} et~al.,}{{Fredslund
  Andersen} et~al.}{2019}]{andersen19}
{Fredslund Andersen} M.,  et~al., 2019, \mn@doi [\aap]
  {10.1051/0004-6361/201935175}, \href
  {https://ui.adsabs.harvard.edu/abs/2019A&A...623L...9F} {623, L9}

\bibitem[\protect\citeauthoryear{{Fr{\"o}hlich} et~al.,}{{Fr{\"o}hlich}
  et~al.}{1995}]{froehlich95}
{Fr{\"o}hlich} C.,  et~al., 1995, \mn@doi [\solphys] {10.1007/BF00733428},
  \href {https://ui.adsabs.harvard.edu/abs/1995SoPh..162..101F} {162, 101}

\bibitem[\protect\citeauthoryear{{Frohlich} et~al.,}{{Frohlich}
  et~al.}{1997}]{froehlich97}
{Frohlich} C.,  et~al., 1997, \mn@doi [\solphys] {10.1023/A:1004969622753},
  \href {https://ui.adsabs.harvard.edu/abs/1997SoPh..170....1F} {170, 1}

\bibitem[\protect\citeauthoryear{{Gabriel} et~al.,}{{Gabriel}
  et~al.}{1995}]{gabriel95}
{Gabriel} A.~H.,  et~al., 1995, \mn@doi [\solphys] {10.1007/BF00733427}, \href
  {https://ui.adsabs.harvard.edu/abs/1995SoPh..162...61G} {162, 61}

\bibitem[\protect\citeauthoryear{{Garc{\'\i}a} et~al.,}{{Garc{\'\i}a}
  et~al.}{2005}]{garcia05}
{Garc{\'\i}a} R.~A.,  et~al., 2005, \mn@doi [\aap]
  {10.1051/0004-6361:20052779}, \href
  {https://ui.adsabs.harvard.edu/abs/2005A&A...442..385G} {442, 385}

\bibitem[\protect\citeauthoryear{{Gaulme} et~al.,}{{Gaulme}
  et~al.}{2016}]{gaulme16}
{Gaulme} P.,  et~al., 2016, \mn@doi [\apjl] {10.3847/2041-8213/833/1/L13},
  \href {https://ui.adsabs.harvard.edu/abs/2016ApJ...833L..13G} {833, L13}

\bibitem[\protect\citeauthoryear{{Hale}, {Howe}, {Chaplin}, {Davies}  \&
  {Elsworth}}{{Hale} et~al.}{2016}]{hale16}
{Hale} S.~J.,  {Howe} R.,  {Chaplin} W.~J.,  {Davies} G.~R.,   {Elsworth}
  Y.~P.,  2016, \mn@doi [\solphys] {10.1007/s11207-015-0810-0}, \href
  {http://adsabs.harvard.edu/abs/2016SoPh..291....1H} {291, 1}

\bibitem[\protect\citeauthoryear{{Harvey} et~al.,}{{Harvey}
  et~al.}{1996}]{harvey96}
{Harvey} J.~W.,  et~al., 1996, Science, \href
  {http://adsabs.harvard.edu/cgi-bin/nph-bib_query?bibcode=1996Sci...272.1284H&db_key=AST}
  {272, 1284}

\bibitem[\protect\citeauthoryear{{Howe}, {Davies}, {Chaplin}, {Elsworth}  \&
  {Hale}}{{Howe} et~al.}{2015}]{howe15}
{Howe} R.,  {Davies} G.~R.,  {Chaplin} W.~J.,  {Elsworth} Y.~P.,   {Hale}
  S.~J.,  2015, \mn@doi [\mnras] {10.1093/mnras/stv2210}, \href
  {https://ui.adsabs.harvard.edu/abs/2015MNRAS.454.4120H} {454, 4120}

\bibitem[\protect\citeauthoryear{{Huber}, {Stello}, {Bedding}, {Chaplin},
  {Arentoft}, {Quirion}  \& {Kjeldsen}}{{Huber} et~al.}{2009}]{huber09}
{Huber} D.,  {Stello} D.,  {Bedding} T.~R.,  {Chaplin} W.~J.,  {Arentoft} T.,
  {Quirion} P.~O.,   {Kjeldsen} H.,  2009, Communications in Asteroseismology,
  \href {https://ui.adsabs.harvard.edu/abs/2009CoAst.160...74H} {160, 74}

\bibitem[\protect\citeauthoryear{{Huber} et~al.,}{{Huber}
  et~al.}{2011}]{huber11}
{Huber} D.,  et~al., 2011, \mn@doi [\apj] {10.1088/0004-637X/743/2/143}, \href
  {https://ui.adsabs.harvard.edu/abs/2011ApJ...743..143H} {743, 143}

\bibitem[\protect\citeauthoryear{{Jim{\'e}nez-Reyes}, {Chaplin}, {Elsworth},
  {Garc{\'\i}a}, {Howe}, {Socas-Navarro}  \& {Toutain}}{{Jim{\'e}nez-Reyes}
  et~al.}{2007}]{jimenezreyes07}
{Jim{\'e}nez-Reyes} S.~J.,  {Chaplin} W.~J.,  {Elsworth} Y.,  {Garc{\'\i}a}
  R.~A.,  {Howe} R.,  {Socas-Navarro} H.,   {Toutain} T.,  2007, \mn@doi [\apj]
  {10.1086/509700}, \href
  {https://ui.adsabs.harvard.edu/abs/2007ApJ...654.1135J} {654, 1135}

\bibitem[\protect\citeauthoryear{{Jim{\'e}nez}, {Garc{\'\i}a}  \&
  {Pall{\'e}}}{{Jim{\'e}nez} et~al.}{2011}]{jimenez11}
{Jim{\'e}nez} A.,  {Garc{\'\i}a} R.~A.,   {Pall{\'e}} P.~L.,  2011, \mn@doi
  [\apj] {10.1088/0004-637X/743/2/99}, \href
  {https://ui.adsabs.harvard.edu/abs/2011ApJ...743...99J} {743, 99}

\bibitem[\protect\citeauthoryear{Jones, Oliphant, Peterson  et~al.}{Jones
  et~al.}{2001}]{scipy}
Jones E.,  Oliphant T.,  Peterson P.,   et~al., 2001, {SciPy}: Open source
  scientific tools for {Python}, \url {http://www.scipy.org/}

\bibitem[\protect\citeauthoryear{{Kallinger} et~al.,}{{Kallinger}
  et~al.}{2010}]{kallinger10}
{Kallinger} T.,  et~al., 2010, \mn@doi [\aap] {10.1051/0004-6361/201015263},
  \href {https://ui.adsabs.harvard.edu/abs/2010A&A...522A...1K} {522, A1}

\bibitem[\protect\citeauthoryear{{Kallinger} et~al.,}{{Kallinger}
  et~al.}{2014}]{kallinger14}
{Kallinger} T.,  et~al., 2014, \mn@doi [\aap] {10.1051/0004-6361/201424313},
  \href {https://ui.adsabs.harvard.edu/abs/2014A&A...570A..41K} {570, A41}

\bibitem[\protect\citeauthoryear{{Kiefer}, {Komm}, {Hill}, {Broomhall}  \&
  {Roth}}{{Kiefer} et~al.}{2018}]{kiefer18}
{Kiefer} R.,  {Komm} R.,  {Hill} F.,  {Broomhall} A.-M.,   {Roth} M.,  2018,
  \mn@doi [\solphys] {10.1007/s11207-018-1370-x}, \href
  {https://ui.adsabs.harvard.edu/abs/2018SoPh..293..151K} {293, 151}

\bibitem[\protect\citeauthoryear{{Kjeldsen} \& {Bedding}}{{Kjeldsen} \&
  {Bedding}}{1995}]{kjeldsen95}
{Kjeldsen} H.,  {Bedding} T.~R.,  1995, \aap, \href
  {https://ui.adsabs.harvard.edu/abs/1995A&A...293...87K} {293, 87}

\bibitem[\protect\citeauthoryear{{Lund} et~al.,}{{Lund} et~al.}{2017}]{lund17}
{Lund} M.~N.,  et~al., 2017, \mn@doi [\apj] {10.3847/1538-4357/835/2/172},
  \href {https://ui.adsabs.harvard.edu/abs/2017ApJ...835..172L} {835, 172}

\bibitem[\protect\citeauthoryear{{Mosser}, {Samadi}  \& {Belkacem}}{{Mosser}
  et~al.}{2013}]{mosser13}
{Mosser} B.,  {Samadi} R.,   {Belkacem} K.,  2013, in {Cambresy} L.,  {Martins}
  F.,  {Nuss} E.,   {Palacios} A.,  eds, SF2A-2013: Proceedings of the Annual
  meeting of the French Society of Astronomy and Astrophysics. pp 25--36
  (\mn@eprint {arXiv} {1310.4748})

\bibitem[\protect\citeauthoryear{{Newville} et~al.,}{{Newville}
  et~al.}{2018}]{LMFIT2018}
{Newville} M.,  et~al., 2018, {lmfit-py 0.9.12},
  \mn@doi{10.5281/zenodo.1699739}

\bibitem[\protect\citeauthoryear{{Nielsen}, {Schunker}, {Gizon}, {Schou}  \&
  {Ball}}{{Nielsen} et~al.}{2017}]{nielsen17}
{Nielsen} M.~B.,  {Schunker} H.,  {Gizon} L.,  {Schou} J.,   {Ball} W.~H.,
  2017, \mn@doi [\aap] {10.1051/0004-6361/201730896}, \href
  {https://ui.adsabs.harvard.edu/abs/2017A&A...603A...6N} {603, A6}

\bibitem[\protect\citeauthoryear{Oliphant}{Oliphant}{2006}]{oliphant2006guide}
Oliphant T.~E.,  2006, A guide to NumPy

\bibitem[\protect\citeauthoryear{{Price-Whelan} et~al.,}{{Price-Whelan}
  et~al.}{2018}]{astropy:2018}
{Price-Whelan} A.~M.,  et~al., 2018, \mn@doi [\aj] {10.3847/1538-3881/aabc4f},
  \href {https://ui.adsabs.harvard.edu/#abs/2018AJ....156..123T} {156, 123}

\bibitem[\protect\citeauthoryear{Stancik \& Brauns}{Stancik \&
  Brauns}{2008}]{STANCIK200866}
Stancik A.~L.,  Brauns E.~B.,  2008, \mn@doi [Vibrational Spectroscopy]
  {https://doi.org/10.1016/j.vibspec.2008.02.009}, 47, 66

\bibitem[\protect\citeauthoryear{Tapping}{Tapping}{2013}]{Tapping2013}
Tapping K.~F.,  2013, \mn@doi [Space Weather] {10.1002/swe.20064}, 11, 394

\bibitem[\protect\citeauthoryear{{Viani}, {Basu}, {Chaplin}, {Davies}  \&
  {Elsworth}}{{Viani} et~al.}{2017}]{viani17}
{Viani} L.~S.,  {Basu} S.,  {Chaplin} W.~J.,  {Davies} G.~R.,   {Elsworth} Y.,
  2017, \mn@doi [\apj] {10.3847/1538-4357/aa729c}, \href
  {https://ui.adsabs.harvard.edu/abs/2017ApJ...843...11V} {843, 11}

\makeatother
\end{thebibliography}

\bsp	
\label{lastpage}
\end{document}